\newcommand {\ignore}[1]{}

\newcommand{\noi}{\noindent}
\newcommand{\bc}{\begin{center}}
\newcommand{\ec}{\end{center}}

\def\ifmath#1{\relax\ifmmode #1\else $#1$\fi}
\def\3quarter{{\textstyle{3 \over 4}}}

\def\ra{\rightarrow}

\def\lf{\leaders\hbox to 1em{\hss.\hss}\hfill}

\def\21{$SU(2) \ot U(1)$}
\def\321{$SU(3) \ot SU(2) \ot U(1)$}
\def\ne{\hbox{$\nu_e$ }}
\def\nm{\hbox{$\nu_\mu$ }}
\def\nt{\hbox{$\nu_\tau$ }}

\def\O{\hbox{$\cal O$ }}
\def\L{\hbox{$\cal L$ }}
\def\mne{\hbox{$m_{\nu_e}$ }}
\def\mnm{\hbox{$m_{\nu_\mu}$ }}
\def\mnt{\hbox{$m_{\nu_\tau}$ }}

\def\etal{\hbox{\it et al., }}

\def\gau{\hbox{gauge }}

\def\neu{\hbox{neutrino }}

\def\neus{\hbox{neutrinos }}

\def\eq#1{{eq. (\ref{#1})}}

\def\VEV#1{\left\langle #1\right\rangle}

\def\ltap{\raisebox{-.4ex}{\rlap{$\sim$}} \raisebox{.4ex}{$<$}}
\def\gtap{\raisebox{-.4ex}{\rlap{$\sim$}} \raisebox{.4ex}{$>$}}
\def\lsim{\raise0.3ex\hbox{$\;<$\kern-0.75em\raise-1.1ex\hbox{$\sim\;$}}}
\def\gsim{\raise0.3ex\hbox{$\;>$\kern-0.75em\raise-1.1ex\hbox{$\sim\;$}}}

\def\bel{\begin{letter}}
\def\eel{\end{letter}}
\def\beq{\begin{equation}}
\def\eeq{\end{equation}}
\def\bef{\begin{figure}}
\def\eef{\end{figure}}
\def\bet{\begin{table}}
\def\eet{\end{table}}
\def\bea{\begin{eqnarray}}
\def\ba{\begin{array}}
\def\ea{\end{array}}
\def\bi{\begin{itemize}}
\def\ei{\end{itemize}}
\def\ben{\begin{enumerate}}
\def\een{\end{enumerate}}
\def\ra{\rightarrow}
\def\ot{\otimes}

\def\eea{\end{eqnarray}}
\def\apj#1#2#3{          {\it Astrophys. J. }{\bf #1} (19#2) #3}

\def\nat#1#2#3{          {\it Nature }{\bf #1} (19#2) #3}
\def\nps#1#2#3{          {\it Nucl. Phys. B (Proc. Suppl.) }
                         {\bf #1} (19#2) #3} 
\def\np#1#2#3{           {\it Nucl. Phys. }{\bf #1} (19#2) #3}
\def\pl#1#2#3{           {\it Phys. Lett. }{\bf #1} (19#2) #3}
\def\pr#1#2#3{           {\it Phys. Rev. }{\bf #1} (19#2) #3}
\def\prep#1#2#3{         {\it Phys. Rep. }{\bf #1} (19#2) #3}
\def\prl#1#2#3{          {\it Phys. Rev. Lett. }{\bf #1} (19#2) #3}

\def\n.c.#1#2#3{         {\it Nuovo Cim. }{\bf #1} (19#2) #3}
\def\r.n.c.#1#2#3{       {\it Riv. del Nuovo Cim. }{\bf #1} (19#2) #3}

\def\zetfpr#1#2#3{         {\it Z. Eksp. Teor. Fiz. Pisma. Red. }{\bf #1} (19#2) #3}

\def\mpl#1#2#3{          {\it Mod. Phys. Lett. }{\bf #1} (19#2) #3}

\def\ppnp#1#2#3{           {\it Prog. Part. Nucl. Phys. }{\bf #1} (19#2) #3}

\def\opc{\hbox{{\sl op. cit.} }}

\relax
\documentstyle[12pt]{article}
\def\lsim{\:\raisebox{-0.5ex}{$\stackrel{\textstyle<}{\sim}$}\:}
\def\gsim{\:\raisebox{-0.5ex}{$\stackrel{\textstyle>}{\sim}$}\:}
\begin{document}
\begin{titlepage}
\begin{center}
\hfill{FTUV/94-51}\\
\hfill{IFIC/94-46}\\
\vskip 0.3cm
{\Large \bf LEFT-RIGHT SYMMETRY AND NEUTRINO STABILITY}\\
\vskip 1.0cm
{\large \bf EUGENI Kh. AKHMEDOV}
\footnote{On leave from National Research Center Kurchatov Institute, 
123182 Moscow, Russia}
\footnote{Present address: SISSA, Via Beirut 2-4, 34014, Trieste, Italy}
\vskip .5cm
{\large \bf ANJAN S. JOSHIPURA}
\footnote{Permanent address: Theory Group,
 Physical Research Lab., Ahmedabad, India}
\vskip .5cm
{\large \bf STEFANO RANFONE}
\footnote{E-mail 16444::RANFONE}\\
\vskip .5cm
and\\
\vskip .5cm
{\large \bf JOS\'E W. F. VALLE}
\footnote{E-mail VALLE at vm.ci.uv.es or 16444::VALLE}\\
\vskip .5cm
{\it Instituto de F\'{\i}sica Corpuscular - C.S.I.C.\\
Departament de F\'{\i}sica Te\`orica, Universitat de Val\`encia\\
46100 Burjassot, Val\`encia, SPAIN}\\
\end{center}

\begin{abstract}
\baselineskip=12pt
{
We consider a left-right symmetric model in which \neus acquire 
mass due to the spontaneous violation of both the gauged $B-L$ and 
a global $U(1)$ symmetry 
broken by the vacuum expectation value (VEV) of a \gau singlet scalar 
boson $\VEV{\sigma}$.
For suitable choices of $\VEV{\sigma}$ consistent
with all laboratory and astrophysical observations
\neus will be unstable against majoron emission.
All \neu masses in the keV to MeV range are possible,
since the expected \neu decay lifetimes can be short
enough to dilute their relic density below the cosmologically
required level. A wide variety of possible new phenomena, 
associated to the presence of left-right symmetry and/or the
global symmetry at the TeV scale, could
therefore be observable, without conflict with cosmology. 
The latter includes the possibility of invisibly decaying higgs 
bosons, which can be searched at LEP, NLC and LHC. }
\end{abstract}

\end{titlepage}

\section{Introduction}

One of the most attractive extensions of the standard electroweak
theory is based on the gauge group
$G_{LR} \equiv SU (2)_L \otimes SU (2)_R \otimes U (1)_{B-L}$
\cite{LR1,LR2}. Apart from offering a possibility of understanding
parity violation on the same footing as that of the gauge symmetry,
these models incorporate naturally small neutrino masses.
The magnitude of these masses is related in these theories
to the scale at which the $SU(2)_R$ symmetry gets broken.
If this breaking occurs at a low ($\sim $ 10 TeV) scale,
then the neutrino masses are expected near their present
laboratory limits, at least for sizeable values of the
Dirac neutrino masses.

Such high values of the neutrino masses may be more
than a theoretical curiosity, they may have quite
important implications. For example, a tau neutrino with a mass
in the MeV range is an interesting possibility first
because such a neutrino is within the range of the
detectability, for example at a tau-charm factory
\cite{jj}. On the other hand, if such neutrino
decays  before the matter dominance epoch, its decay
products could then add energy to the radiation thereby
delaying the time at which the matter and radiation
contributions to the energy density of the universe
become equal. This would reduce density fluctuations
on smaller scales \cite{latedecay} purely within the
framework of the standard cold dark matter model \cite{cdm},
and could reconcile the large scale fluctuations observed by
COBE \cite{cobe} with the earlier observations such as those
of IRAS \cite{iras} on the fluctuations at smaller scales
\cite{ma1}.

It is well known that, if stable, neutrinos would contribute
too much to the energy density of the universe
if their mass lies in the range \cite{RHOCRIT}
\beq
60 \:{\rm eV}\: \ltap \: m_{\nu} \: \ltap \;{\rm few \; GeV}
\label{RHO1}
\eeq
Thus the interesting possibility of heavy neutrino masses
can be consistent with cosmology only if there are new
neutrino decay and/or annihilation channels absent in the 
standard model. 
Many neutrino decay modes have been suggested but
all are quite unlikely to breach the forbidden range
given above \cite{fae}. For example, neutrino radiative
decay modes $\nu' \rightarrow \nu + \gamma$ and
$\nu' \rightarrow \nu + \gamma+ \gamma$ are disfavored,
because they have a very long lifetime \cite{PAL.WOLF}.
Moreover, such {\sl visible} decays are very constrained by
astrophysics \cite{SATO.SARKAR} as well as laboratory
searches \cite{OBE}. What one needs are {\sl invisible}
decays such as $\nu' \rightarrow 3 \nu$. It was noted long
ago that such decays take place in models where isodoublet
and isosinglet mass terms coexist, due to the peculiar
structure of the neutral current in these models \cite{2227,774}
and this is the case in the left-right models. In contrast to the
visible decays, these are almost unconstrained. However, in the
simplest models of the seesaw type even if \neu masses are close
to their laboratory limits, the expected lifetimes tend to be
too long to allow for sufficient redshift of the heavy
neutrino decay products, and thus forbidden by
cosmology \cite{774}. Moreover, for
$m_{\nu'} \gsim 1\:{\rm MeV}$ the $3\nu$ decay would also be
accompanied by the {\sl visible} channel
$\nu' \rightarrow e^+ e^- \nu$. This would, in turn
suggest a $\gamma$-ray burst from a supernova explosion,
the photons arising from subsequent annihilation and/or
bremsstrahlung processes. The non-observation of such a
burst from SN1987 disfavours this possibility \cite{DAR}.

Although not possible in the simplest models \cite{774},
fast invisible \neu decays can, under certain circumstances,
naturally occur in many models where \neu masses are induced
from the spontaneous violation of a global $B-L$ symmetry
\cite{V,CON,fae}
\beq
\nu' \rightarrow \nu + J \:,
\label{NUJ}
\eeq
where here $J$ denotes the massless Nambu-Goldstone
boson, called majoron \cite{CMP}, which follows from
the spontaneous nature of lepton number violation.
These decays could have important implications
in cosmology and astrophysics \cite{fae}.

Unfortunately this possibility does not arise naturally
in the left-right symmetric framework since the global
symmetry associated with the conventional majoron is gauged
in this case. Thus in order to obtain majoron one needs to
impose an additional symmetry which is different from the
$B-L$ symmetry, but which nevertheless plays a role in
generating the neutrino masses and decays.

In this paper we propose a variant of the left-right symmetric
model with an additional spontaneously broken $U(1)$ global
symmetry, acting nontrivially on some new isosinglet leptons
which mix with the ordinary neutrinos.  Thus we extend the fermion 
sector in order to accommodate
the required global symmetry whose spontaneous breaking will yield the
majoron. This allows us to incorporate the idea of
invisibly decaying neutrinos in the framework of a theory with gauged
$B-L$. The additional singlet fermions used in our model may arise in
various attempts to unify quarks and leptons in a superstring
framework \cite{SST1}.

Majoron decays of neutrinos in the left-right symmetric model 
has also been considered in ref. \cite{RabiPal}. However,
our model has a more economic higgs sector and makes
use of a different fermion content. Therefore, in a 
sense, it is complementary to that of \cite{RabiPal}. 

In fact, our model is a left-right embedding of a
previously suggested model \cite{CON} but has some noticeable
differences which we study.
We investigate the issue of \neu stability in this model
and demonstrate that, for reasonable choices of the breaking
scales, $v_R \gsim \omega$ ($v_R$ is the $B-L$ and parity 
breaking scale
while $\omega \equiv \VEV{\sigma}$ characterizes the breaking
of the global $U(1)_G$ symmetry), the \neu decay amplitude for
the majoron decay mode of \eq{NUJ} is of order $(m_D/M)^4$ 
where $M = g_5 v_R$ with $g_5$ being the appropriate Yukawa
coupling. This is in full agreement with previous studies within
the \21 theory \cite{774,V}. Nevertheless, for appropriate
choices of parameters, this simplest model yields majoron
emission \nt decay lifetimes which can be fast enough to
dilute the relic \nt density to acceptable levels for all
values of the \nt mass. For most typical parameter
choices, the \nm is light enough as to lie outside the
range in \eq{RHO1} and be stable, as required in
order to be hot dark matter.

We also propose a very simple variant of this left-right
symmetric model where the global $U(1)_G$ symmetry is of the
horizontal type, as originally used in ref. \cite{V}. This
substantially enhances the \neu decay amplitude for
the majoron decay mode of \eq{NUJ} to order $(m_D/M)^2$.
In this case the majoron is a pure \gau singlet, as in the
original proposal \cite{CMP}, and therefore both scales
$\VEV{\sigma}$ and $v_R$ may be chosen to be at the TeV
scale quite naturally. This opens up a very wide
phenomenological potential for left-right
extensions of the standard electroweak theory,
free of cosmological problems.

\section{The simplest model}

We consider a model based on the \gau group
$$
G_{LR} \equiv SU (2)_L \otimes SU (2)_R \otimes U (1)_{B - L}
$$
in which an extra $U(1)_G$ global symmetry is postulated.
The matter and higgs boson representation content is specified
in table 1. In addition to the conventional quarks and leptons,
there is a \gau singlet fermion in each generation
\footnote{Although the number of such singlets is arbitrary,
since they do not carry any anomaly, we add just one such
lepton in each generation, while keeping the quark sector as
the standard one. Further extensions can be made, as recently
discussed in ref. \cite{Ma}. }.
These extra leptons might arise in superstring
models \cite{SST1}. They have also been discussed
in an early paper of Wyler and Wolfenstein \cite{WYLER}.
We will not use the more conventional triplet higgs scalars,
which are absent in many of these string models. Instead
we will substitute them by the doublets $\chi_{L}$ and $\chi_{R}$.
This could in fact play an important role in
unifying this model in $SO(10)$, while keeping left-right
symmetry unbroken down to the TeV scale \cite{Ma}.

The Yukawa interactions allowed by the $G_{LR} \otimes U(1)_G$
symmetry are given as
\beq
\begin{array}{cc}
- \L_Y & = g_1 \bar{Q}_L \phi Q_R + g_2 \bar{Q}_L \tilde{\phi} Q_R
+ g_3 \bar{\psi}_L \, \phi \,  \psi_R +
g_4 \bar{\psi}_L \tilde{\phi} \psi_{R} + \\
& g_5 [\bar{\psi}_{L} \chi_L S^c_R + \bar{\psi}_R \chi_R S_L ] +
g_6 \bar{S}_L S^c_R \sigma + h.c.
\end{array}
\eeq
where $g_i$ are matrices in generation space and
$\tilde{\phi} = \tau_2 \phi^* \tau_2 $ denotes the
conjugate of $\phi$. This Lagrangean is invariant
under parity operation $Q_L \leftrightarrow Q_R$, $\psi_L \leftrightarrow 
\psi_R$, $S_L \leftrightarrow S^c_R$, $\phi \leftrightarrow \phi^\dagger$ 
and $\chi_L \leftrightarrow \chi_R$.

The symmetry breaking pattern is specified by the
following scalar boson VEVs (assumed real):
\beq
\left\{
\begin{array}{lll}
\VEV{\phi} =  \left( \begin{array}{cc}
k & 0 \\
0 & k'
\end{array} \right) \:; & \VEV{\chi_L^0} = v_L \:; & \VEV{\chi_R^0} = v_R\,;\\
[2ex]
\VEV{\tilde{\phi} } = \left(
\begin{array}{cc}
k' & 0 \\
0 & k
\end{array} \right) \:; & \VEV{\sigma } = \omega
\end{array} \right.
\eeq
The spontaneous violation of the global $U(1)_G$ symmetry generates
a physical majoron whose profile in the limit $V^2 \ll v_R^2$
is specified as
\beq
\label{profi}
 J   = ( \omega^2 + \frac{v_L^2 v^2}{V^2} )^{- 1/2}    \{ \omega
 \sigma_I  + \frac{v_L v}{V^2} [ v  \chi_L^I
 - \frac{v_L}{v} (k  \phi_2  - k' \phi_4 )] \}
\eeq
where  $\sigma_I$, $\chi_L^I$, $\phi_2$ and $\phi_4$
denote the imaginary parts of the neutral fields
in $\sigma$, $\chi_L$ and the bidoublet $\phi$.
Here we have also defined the VEVs as
$v^2 \equiv k^2 + k'^2$ and $V^2 \equiv v^2 + v_L^2$.

Note that the majoron has no component along the imaginary
part of $\chi_R$ despite the fact that $\chi_R$ is
nontrivial under the global symmetry. Clearly,
as it must, the majoron is orthogonal to the Goldstone
bosons eaten-up by the $Z$ and the new heavier neutral
gauge boson present in the model. The latter
acquires mass at the larger scale $v_R$.

The various scales appearing in \eq{profi} are not
arbitrary. First of all, note that the minimization
of the scalar potential dictates the consistency relation
\cite{LR2}
\beq
\label{vevseesaw}
\frac{v_R}{v} \sim \lambda \frac{\omega }{v_L}
\eeq
For $\lambda \sim 1$ the singlet VEV is necessarily
larger than $v_L$ i.e. $ v_L \ll \omega$ and, as a result,
the majoron is mostly singlet and the invisible decay  of
the $Z$ to the majoron is enormously suppressed, unlike
in the purely doublet or triplet majoron schemes.

On the other hand, in order that majoron emission
does not overcontribute to stellar energy loss 
one needs to require \cite{DEAR}
\beq
\label{red}
\frac{v_L^2}{\sqrt{2} \omega V^2} \lsim 10^{-9} \rm{GeV^{-1}}\;.
\eeq
One sees that \eq{vevseesaw} and \eq{red} allow for
the existence of right-handed weak interactions at
accessible levels, provided $v_L$ is sufficiently
small, i.e. \O($\lsim$ 100 keV).

Note, however that the astrophysical bound in \eq{red}
is hard to reconcile with the low-scale right-handed weak
interactions in the case where $v_L \sim v$. For example,
$v_L \sim 1$ GeV would require $v_R \; \gtap \; 10^{7}$ GeV, 
$\omega \; \gtap \; 10^{5}$ GeV 
with $v^2 \equiv k^2 + k'^2$ fixed by the masses
of $W$ and $Z$ bosons. As we will show later, it is
possible to avoid this bound altogether in a
simple variant of this model (see below).

\section{Neutrino Masses and Majoron Couplings}

Once all \gau and global symmetries get broken
a mass term is generated for the electrically
neutral leptons, of the form
$$
\frac{1}{2} \left( \bar{\Psi}_L M_\nu \Psi_R + h.c. \right)\;,
$$
where $\Psi_L = (\nu_L, \nu^c_L, S_L)$ and
$\Psi_R = (\nu_{R}^{c}, \nu_R, S^c_R)$. It
may be written in block form as
\beq
\label{MAT}
\quad \quad
M_{\nu} = \left(
\begin{array}{c}
\end{array}
\begin{array}{ccc}
0 & m_D & \beta\\
m_D^T & 0 & M^*\\
\beta^{T} & M^{\dagger} & \mu
\end{array} \right)\;,
\eeq
where the various entries are specified as
\beq
\beta = g_5 v_L \, , \quad M = g_5 v_R \, \quad  m_D = g_3 k +
g_4 k' \, , \quad  \mu = 2 g_6 \omega .
\eeq
Here the matrix $m_{D}$ is the Dirac mass term
determined by the standard higgs bi-doublet VEV
$\VEV{\phi}$ responsible for quark and charged
lepton masses, $\beta$ and $M$ are $G$ and $B-L$ violating
mass terms determined by $v_L$ and $v_R$, while
$\mu$ is a gauge singlet $G$-violating mass,
proportional to the VEV of the gauge singlet
higgs scalar $\sigma$ carrying 2 units of $G$ charge.
Note the zeroes in the first two diagonal entries.
They arise because there are no higgs fields to provide
the usual Majorana mass terms \cite{SST1} which
would be required in the seesaw mechanism \cite{GRS,LR2,774}.

In order to determine the light \neu masses and
majoron couplings we will work in the {\sl seesaw}
approximation, which
we define as $M, \mu \gg m_D, \beta$. In this case the
mass matrix in \eq{MAT} can be brought to block
diagonal form via a transformation $U$ (with $U^{\dagger} U = U
U^{\dagger} = 1$), 
\beq
\label{M1}
\hat{M}_\nu \equiv \, U \, M_{\nu} \, U^T =
\left( \begin{array}{cc}
\hat{M}_{1} & 0 \\
0 & \hat{M}_2
\end{array}
\right)\;,
\eeq
where
\beq
\label{M2}
\begin{array}{cc}
\hat{M}_1 &  =  - (m_D, \: \beta) \left( \begin{array}{cc}
0 & M^* \\
M^\dagger & \mu
\end{array}
\right)^{-1}
\left (  \begin{array}{c}
m_D^T\\
\beta^T \end{array}
\right) = \\[2ex]
& =  \epsilon \mu \epsilon^T - ( \beta \epsilon^T +
\epsilon \beta^T ), \qquad \epsilon \equiv m_D M^{\dagger -1}
\end{array}
\eeq
denotes the
effective light \neu mass matrix, determining
the masses of \ne, \nm and \nt. Notice that the
light \neu masses are generated due to the interplay of the 
violation of the global $G$ as well as the gauged $B-L$ symmetry. 
Due to the relation in \eq{vevseesaw} the two contributions to the
\neu masses in the last line of \eq{M2} will be typically comparable.

The heavy sector is characterized by a 6 $\times$ 6
mass matrix given as
$$
\hat{M}_{2} \simeq M_2 \equiv \left(
\begin{array}{cc}
0 & M^*\\
M^{\dagger} & \mu
\end{array}
\right)\,.
$$
Finally, the matrix $\hat{M}_{\nu}$ is further diagonalized
by a block diagonal unitary matrix T
\beq
T \hat{M}_\nu T^T = M_{diag} = {\rm diag} (m_{1}, ... m_{9})\,.
\eeq
which can be written as
$$
T = \left( \begin{array}{cc}
V_1 & 0\\
0   & V_2
\end{array} \right) ,
$$
The total diagonalizing matrix A can then be written as
\beq
\label{A}
A = T U = \left(
\begin{array}{cc}
V_1 (1 - \frac{1}{2}  \rho \rho^\dagger  ) & - V_1 \rho\\
V_2 \rho^\dagger & V_2 (1 - \frac{1}{2} \rho^\dagger  \rho)
\end{array}
\right) \, + \O (\rho^3 )\,,
\eeq
where $V_1$ and $V_2$ are the matrices that diagonalize the
light and heavy \neu mass matrices respectively, and
\beq
\label{rho}
\rho \equiv (m_{D}, \beta ) M_{2}^{-1} =
( - (\epsilon \mu - \beta) M^{*-1} , \epsilon)\,.
\eeq
Note that $\rho \ra 0$ as $M \ra \infty$. This parameter
plays the same role in the present model as the expansion
parameter $\epsilon$ introduced in the \21 context in
ref. \cite{774}. The relation between weak and mass
eigenstates may then be written as
$$
(\nu^c_R, \nu_R, S^c_R )^T = A^T (\nu^{c}_R{}', \nu_R{}', 
S^{c}_R{}')^T\;,
$$
where the prime refers to the mass eigenstate basis.
The majoron-neutrino interaction Lagrangean, obtainable using eqs.(3) and (5),
may be written in terms of weak eigenstates as
\beq
\label{Jnn}
\L_J = \frac{i J}{\sqrt{2}} 
\frac{1} {\sqrt{\omega^2 + \left(\frac{v_L v}{V} \right)^2}}  
\left\{ 
\frac{1}{2} \bar{S}_L \mu S^{c}_R + 
\left( \frac{v}{V} \right)^2 \, \bar{\nu}_L \beta S^c_R -
\left( \frac{v_L}{V} \right)^2 \, \bar{\nu}_L m_D \nu_R \, \right\} + h.c.
\eeq

\section{Neutrino Decays and Cosmology}

Relic \neus will overcontribute to the present-day
energy density of the universe unless there are
decay and/or annihilation channels.
The cosmological density constraint on the neutrino decay
lifetime for a $m_\nu \lsim 1$ MeV neutrino is given as
\cite{KT,RHOCRIT2}
\beq
m_\nu \left(\frac{\tau_\nu}{t_0}\right)^{1/2} \ltap \; 100\; h^2 \;
\rm{eV} 
\label{RHO2}
\eeq
where $t_0$ and $h$ are the present age of the universe and the
normalized Hubble parameter. The above constraint follows from 
demanding that an adequate redshift of the heavy neutrino decay 
products occurs.

In the present model, even though $B-L$ is a \gau
symmetry, neutrino masses following from \eq{MAT} are
accompanied by the existence of a massless majoron $J$ 
given by \eq{profi}. This will lead to invisible \neu
decays with majoron emission, \eq{NUJ}.

To determine the \neu decay rates we are
interested in those majoron couplings to light \neus
that are nondiagonal in the mass eigenstate basis.
These couplings can be determined by rewriting explicitly
\eq{Jnn} in terms of mass eigenstates. This procedure
is straightforward but subtle. There are, here too, the
same tricky cancellations first discovered in the
context of the standard \21 model in ref. \cite{774}.
The result is that majoron couplings to light \neus
are still diagonal to \O($\epsilon^2$), and therefore
can not induce \neu decay to this order.

In order to see this more clearly and, at the same time,
determine the required nondiagonal couplings we prefer,
instead of directly using \eq{Jnn}, to use a more general
and powerful method based on the use of Noether's
theorem for the global $G$-current. The method was given
in the Sec. VI of ref. \cite{774} and subsequently
used, {\it e.g.}, in the first paper of ref. \cite{V}. It has
the advantages of being simpler and more systematic.

Using it one can easily determine the coupling matrix
of the majoron to the light mass eigenstate \neus in
the present model as
\beq
g_{ab} = \frac{1}{\VEV{\sigma}}[m_a R_{ab} + m_b R_{ba}]\,; \: a \neq b
\label{JCOUP}
\eeq
where $m_a$ denote the light \neu masses and the matrix $R$ is
determined by the three light entries of the 9 $\times$ 9 matrix
\beq
R \approx A^* Q_1 A^T\;,
\eeq
where $Q_1$ is a diagonal matrix related to the $G$ charges of
the leptons $\Psi_L = (\nu_L, \nu^c_L, S_L)$. Since only
the \gau singlet leptons transform under $G$, the matrix $Q_1$
can be written as
\beq
\label{Q1}
Q_1 = \rm{diag} (0, 0, 1)\;.
\eeq
The matrix A was previously defined as that which diagonalizes
the full \neu mass matrix. 
Using \eq{A} we can rewrite the part $R_L$ of the matrix $R$
connecting light neutrinos as the 3 $\times$ 3 matrix
\beq
\label{RL}
R_L = V_1^* \rho^* \hat{Q}_1 \rho^T V_1^T
= V_1^* \epsilon^* \epsilon^T V_1^T\:,
\eeq
where the 6 $\times$ 6 diagonal matrix $\hat{Q_1}$ is
defined as diag(0,0,0,1,1,1). This shows explicitly that
the nondiagonal entries of the majoron coupling matrix in
\eq{JCOUP} arise manifestly at \O($\epsilon^4$), in
agreement with results found in the \21 theory \cite{774}.

In order to get an idea of the expected \neu decay rates in this
model we first make a crude estimate of the magnitude of the
\neu masses following from \eq{M2}. Using \eq{vevseesaw}
and \eq{red} one sees that
\bea
m_{\nu_i} \sim 2 \frac{g_6}{ g_5^2}
\frac{\omega m_{q_i}^2}{v_R^2}
\lsim 2 \frac{g_6}{ g_5^2}
\left(\frac{m_{q_i}}{{\rm GeV}}\right)^2\; \rm{eV}\;.
\eea
In estimating this upper limit 
we have assumed the Dirac \neu masses to be of
the same order as the corresponding up-quark masses.
Assuming a reasonable choice of parameters where
the ratio $(2 g_6\,/ g_5^2 )$ lies in the range
1 to $10^3$ we get
	\mnt$ \lsim $10 keV to 10 MeV,
	\mnm$ \lsim $1 eV to 1 keV
and	\mne$\ \!\!\! \lsim 10^{-4}$ to $10^{-2}$ eV.
Thus the \nm may well be stable, as required in order to be dark
matter, while the \nt is expected to violate the cosmological
limit \eq{RHO1} and has to decay with lifetime obeying \eq{RHO2}.

We now make a simple numerical estimate of the \nt 
lifetime. From \eq{RL} we can parametrize the nondiagonal
coupling responsible for \nt decay to the lighter \neus
plus majoron as
\beq
g_{3a} = \frac{m_3}{\VEV{\sigma}}(g_s + g_c),
\eeq
where $m_3$ denotes \mnt and we have set
\beq
g_s = \frac{1}{2}
[ (\epsilon \epsilon^T)_{22} - (\epsilon \epsilon^T)_{11}] \:
\sin 2\theta_L\;,
\label{GS}
\eeq
\beq
g_c = (\epsilon \epsilon^T)_{12} \:\cos 2\theta_L\;.
\label{GC}
\eeq
In obtaining these formulas we have assumed for simplicity that \nt mixes with only 
one of the two lighter neutrinos, with mixing angle $\theta_L $, 
and that $\epsilon$ is real. 
For small values of $\theta_L $ we may only keep the
second term. Assuming a simple scaling ansatz
$\epsilon \sim m_D/M$ we get
\beq
g_{c} \approx \left(\frac{m_D}{M}\right)^2
\eeq
One can now easily see that the \neu decay lifetime becomes
\beq
\label{taunu}
\tau(\nu_3 \rightarrow \nu + J)
= \frac{16\pi}{g_{3a}^2}\frac{1}{m_{\nu_3}} 
\approx 3 \times 10^{7} \:({\rm keV}/m_3)^{3}
\left(\frac{\VEV{\sigma}}{10^6 {\rm GeV}}\right)^2 \:
\left(\frac{m_D}{M}\right)^{-4} \: \rm{sec}
\eeq
The lifetime in \eq{taunu} can be short enough to
obey the cosmological constraint in \eq{RHO2}.
Indeed, from eqs. (21)--(26) one can readily find the following dependence 
between $\tau(\nu_3)$ and $m_{\nu_3}$ in our model: 

$$ \tau(\nu_3) \approx 1.4\times 10^8 \left(\sqrt{g_6}\frac{\omega}
{{\rm GeV}}\right)^4 \left(\frac{{\rm keV}}{m_{\nu_3}}\right)^5 \;
{\rm sec}\;. $$
This dependence is plotted for two illustrative values of $\sqrt{g_6}\omega$ 
in Fig. 1 alongside with the cosmological bound of \eq{RHO2}.
It can be seen from this figure that for $\sqrt{g_6}\omega=35$ GeV 
cosmology does not constrain the model for $\mnt > 1$ keV. For 
$\sqrt{g_6}\omega=10^3$ GeV cosmological bound excludes values of 
$\mnt$ below 100 keV. Thus in this case cosmological considerations 
provide a lower limit on the $\nt$ mass.

One can rewrite the cosmological constraint \eq{RHO2} in terms of the 
VEVs $\omega$ and $v_R$ instead of $\tau(\nu_3)$ and $m_{\nu_3}$: 
$(\omega/{\rm GeV})^{1/2}(v_R/{\rm GeV})^3 \lsim 6\times 10^{19} A$, where 
$A\equiv \left(\frac{\sqrt{g_6}}{g_5^3}h^2\right)$. The astrophysical 
constraint 
of \eq{red} can be written using \eq{vevseesaw} as $(\omega/{\rm GeV})
(v_R/{\rm GeV})^{-2} < 10^{-9}$. These two constraints are plotted in Fig. 2 
for $A=10^5$. The region below both straight lines illustrates what is allowed
 for this representative choice of parameters. For example, we can see from 
this figure that $\omega$ cannot exceed 
$6\times 10^5$ GeV, corresponding to a value of
$v_R \approx 2.4\times 10^7$ GeV. 

This generalizes to our left-right symmetric model
the results obtained in the analysis of
the question of \neu stability in majoron
models given in ref. \cite{774}. The tau neutrino
is expected to be in the keV to MeV range with a
lifetime that can be as short as 1 sec. 

\section{A Model with Enhanced Neutrino Decays}

We now briefly sketch a variant of the previous model
with exactly the same particle content, but with the
global $U(1)_G$ symmetry $G$ assigned in a nonsequential
way. The model may be seen also as a left-right
symmetric variant of the original horizontal
lepton number models \cite{V}.

The $G$ charges of the lepton doublets
of the first two generations can be assigned as
1 and -1 respectively. Similarly, under the global
$U(1)_G$ symmetry the gauge singlets transform with charges
+1 ($S_{1L}$ and $S_{2R}^c$) and
-1 ($S_{2L}$ and $S_{1R}^c$). Finally the third
generation leptons carry no $G$ charge.

Another important difference with respect to the model discussed
in the previous sections, insofar as the $G$ assignments
of the higgs scalar bosons are concerned, is that
now $\chi_{L}$ and $\chi_{R}$ carry no $G$ charge and
therefore the resulting majoron will be a pure \gau
singlet, as in the original model \cite{CMP}. 
This has a very important phenomenological
implication, namely that one now avoids the
astrophysical constraint of \eq{red}, allowing for very 
low values of the $G$ breaking scale $\VEV{\sigma}$ which may 
naturally lie at the electroweak scale.

The quantum numbers are summarized in Table 2. Since the
same higgs multiplets are used the mass matrix has the
same general structure as in \eq{MAT}. However, as a
result of the horizontal assignment of the global
charges of the leptons, the entries in \eq{MAT}
now have special textures in generation space.

To find these textures, let us first notice that our present 
$G$ charge assignment supports the discrete parity symmetry 
of the model, provided the parity operation for the gauge-singlet 
leptons of the first two generations is modified:  
$S_{Le}\leftrightarrow S^c_{R\mu}$, the rest of the fields 
transforming as before. The $g_5$ term in the Yukawa Lagrangean 
of eq. (3) will now read 
\beq
(g_5)_{1} [\bar{\psi}_{Le} \chi_L S^c_{R\mu} + 
\bar{\psi}_{Re} \chi_R S_{Le} ] +
(g_5)_{2}  [\bar{\psi}_{L\mu} \chi_L S^c_{Re} + 
\bar{\psi}_{R\mu} \chi_R S_{L\mu} ] +
(g_5)_{3}  [\bar{\psi}_{L\tau} \chi_L S^c_{R\tau} + 
\bar{\psi}_{R\tau} \chi_R S_{L\tau} ] 
\eeq
One can now readily find the entries of the neutrino mass matrix 
in \eq{MAT}. They are given by diagonal forms for both $m_D$
and $M$, while the remaining entries
$\beta$ and $\mu$ take on the following forms
\beq
\label{beta}
\beta =
\left(
\begin{array}{ccc}
0 & \times & 0 \\
\times & 0 & 0 \\
0 & 0 & \times
\end{array} \right)
\eeq
and
\beq
\label{mu}
\mu=
\left(
\begin{array}{ccc}
\times & \times & 0 \\
\times & \times & 0 \\
0 & 0 & \times
\end{array} \right)\;,
\eeq
where in the last equation the 12 and 33 entries
are bare masses, allowed by the $G$ symmetry, while
the 11 and 22 are proportional to the VEVs of
$\sigma^*$ and $\sigma$ respectively.

The horizontal nature of the $G$ assignments
removes the additional \O($\epsilon^2$) suppression
in the \neu decay rate. To see this note that now the
matrix $R_L$ of \eq{RL} is replaced by
\beq
\label{RL1}
R_L = V_1^* \hat{Q}_2 V_1^T
\eeq
where $\hat{Q}_2$ is a  3 $\times$ 3 matrix given by
diag (1,-1,0) dictated by the $G$ charge assignments.
Note that in the above equation there is no
$\rho \sim \epsilon$ suppression.

In summary, the main features of this second model are
\ben
\item
The majoron is a pure \gau singlet, allowing for 
the $G$ breaking scale to be as low as the electroweak scale; 
\item
Since \eq{red} need not hold in this model, the
left-right symmetry can be realized at the TeV scale; 
\item
Majoron emission neutrino decay amplitudes are enhanced
to \O($\epsilon^2$). 
\een

The combined effect of the above features is to
provide a tremendous enhancement of the \neu decay
amplitude, leading to a lifetime shorter than \eq{taunu}
by as much as 20 orders of magnitude.

The existence of models such as this opens a very wide
phenomenological potential for left-right extensions of
the standard model, consistent with all cosmological
observations.

\section{Discussion}

We have examined the issue of \neu stability in
a class of left-right symmetric models where \neus
may acquire mass from the spontaneous violation
of both the gauged $B-L$ symmetry and a global $U(1)$ 
symmetry broken by the vacuum expectation value 
of a \gau singlet scalar boson $\sigma$. 
For suitable choices of $\VEV{\sigma}$ consistent
with all laboratory and astrophysical observations
\neus will be unstable against majoron emission.
We have considered two models. In the simplest one
the global symmetry is flavour blind, while in the
second it distinguishes between leptons of
different type. In the first model the tau \neu
may be heavy and unstable against majoron emission,
with decay amplitude of order $(m_D/M)^4$. Despite 
such strong a suppression, neutrino decay rates 
are consistent with the cosmological requirements in 
a wide range of the parameters of the model. 
The parity violation scale can be as low as a few 
TeV if the left-handed doublet higgs VEV $v_L$ is 
$\lsim 100$ keV, but should be $\gsim 10^9$ GeV for
$v_L$ of the order of the electroweak scale. 
In the second model, \neu decay amplitudes are
substantially enhanced by a combined effect
of low values for both global and left-right
symmetry breaking scales $\VEV{\sigma}$ and $v_R$.
These scales may be naturally chosen to be at the TeV scale.
This opens up a very wide phenomenological potential for 
left-right extensions of the standard electroweak theory,
consistent with cosmology. First of all, our models
allow \neu masses in the keV to MeV range, with
potential effects related to \neu masses and
mixing, such as enhanced neutrinoless $\beta \beta$
decay rates. Moreover, they allow for the existence 
of neutral heavy leptons with masses at the weak scale.
If lighter than the Z boson, these may give rise to 
quite striking signatures at LEP \cite{CERN}. 
In addition, there are the potential effects due to
the presence of right-handed weak currents at the
TeV scale, including neutrinoless $\beta \beta$
decays and many other effects. Finally, there may
be effects associated to the global symmetry
violation at the TeV scale, such as the
unusual possibility of an invisibly decaying
higgs boson \cite{JoshipuraValle92} $h \ra JJ$,
which can be searched at future colliders such
as LEP, NLC and LHC \cite{alfonso}.

\noi
{\bf Acknowledgements}
This work was supported by DGICYT under grant number
PB92-0084, by the sabbatical grants SAB93-0090 (E.A.)
and SAB94-0014 (A. J.), and by the EC fellowship grant
number ERBCHBI CT-930726 (S.R.).
We thank G. Senjanovi\'c for fruitful discussions.

\newpage

\begin{tabular}{c|cccc}
 & $ SU (2)_{L} \; \otimes$ & $SU (2)_{R} \; \otimes$  & $U(1)_{B - L} \;
 \otimes$ & $U(1)_G$\\
\hline
$Q_{Li}$ & 2 & 1 & 1/3 & 0\\[0.1cm]
$Q_{Ri}$ & 1 & 2 & 1/3 & 0\\[0.1cm]
$\psi_{Li}$ & 2 & 1 & -1 & 0\\[0.1cm]
$\psi_{Ri}$ & 1 & 2 & -1 & 0\\[0.1cm]
$S_{Li}$ & 1 & 1 & 0 & 1\\[0.1cm]
\hline
$\phi $& 2 &$ 2$ & 0 & 0\\[0.1cm]
$\chi_{L}$ & 2 & 1 & -1 & 1\\[0.1cm]
$\chi_{R}$ & 1 & 2 & -1 & -1\\[0.1cm]
$\sigma$ & 1 & 1 & 0 & 2
\end{tabular}

\noi
Table 1: $SU (2)_L \otimes SU (2)_R \otimes U (1)_{B-L}
\ot U(1)_G$ assignments of the quarks, leptons and higgs
scalars.

\begin{tabular}{c|cccc}
 & $ SU (2)_{L} \; \otimes$ & $SU (2)_{R} \; \otimes$  & $U(1)_{B - L} \;
 \otimes$ & $U(1)_G$\\
\hline
$Q_{Li}$ & 2 & 1 & 1/3 & 0\\[0.1cm]
$Q_{Ri}$ & 1 & 2 & 1/3 & 0\\[0.1cm]
$\psi_{Le}$ & 2 & 1 & -1 & 1\\[0.1cm]
$\psi_{L\mu}$ & 2 & 1 & -1 & -1\\[0.1cm]
$\psi_{L\tau}$ & 2 & 1 & -1 & 0\\[0.1cm]
$\psi_{Re}$ & 1 & 2 & -1 & 1\\[0.1cm]
$\psi_{R\mu}$ & 1 & 2 & -1 & -1\\[0.1cm]
$\psi_{R\tau}$ & 1 & 2 & -1 & 0\\[0.1cm]
$S_{Le}$ & 1 & 1 & 0 & 1\\[0.1cm]
$S_{L\mu}$ & 1 & 1 & 0 & -1\\[0.1cm]
$S_{L\tau}$ & 1 & 1 & 0 & 0\\[0.1cm]
\hline
$\phi $& 2 &$ 2$ & 0 & 0\\[0.1cm]
$\chi_{L}$ & 2 & 1 & -1 & 0\\[0.1cm]
$\chi_{R}$ & 1 & 2 & -1 & 0\\[0.1cm]
$\sigma$ & 1 & 1 & 0 & 2
\end{tabular}

\noi
Table 2: $SU (2)_L \otimes SU (2)_R \otimes U (1)_{B-L}
\ot U(1)_G$ assignments of the quarks, leptons and higgs
scalars in the model of section 5. Notice the nonsequential
assignment of the global charge.

\newpage

{\bf FIGURE CAPTIONS}\\

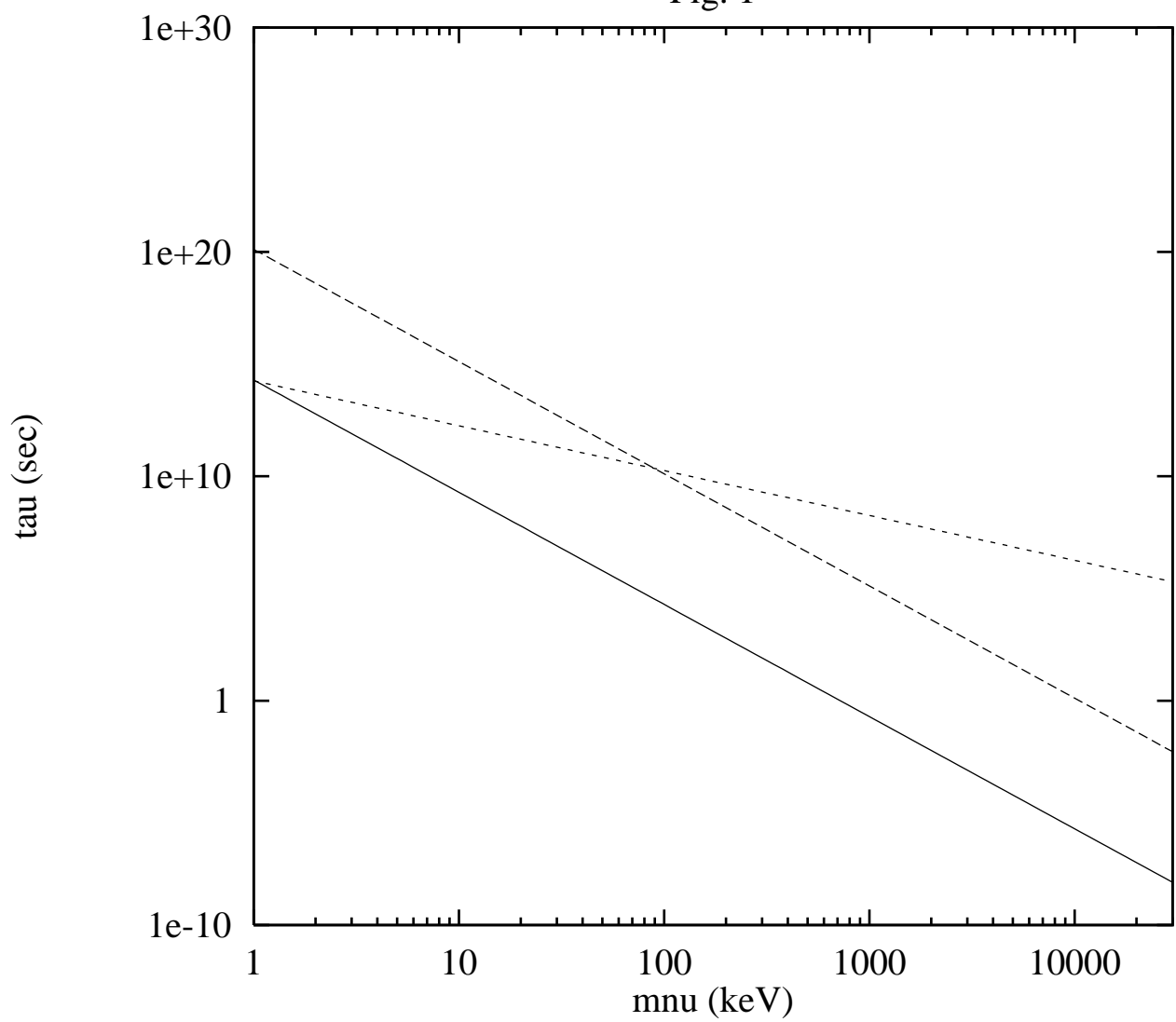
\noindent {\bf Fig. 1}: Typical expectations for the tau neutrino 
lifetime as a function of the neutrino mass in the model of Sec. 2--4. 
The dotted line corresponds
to the cosmological limit of \eq{RHO2} (the region below the line is 
allowed). Solid (dashed) line is the relation between the $\nt$ 
mass and lifetime for two typical choices of the parameter
$\sqrt{g_6} \omega$=35 ($10^3$) GeV. 

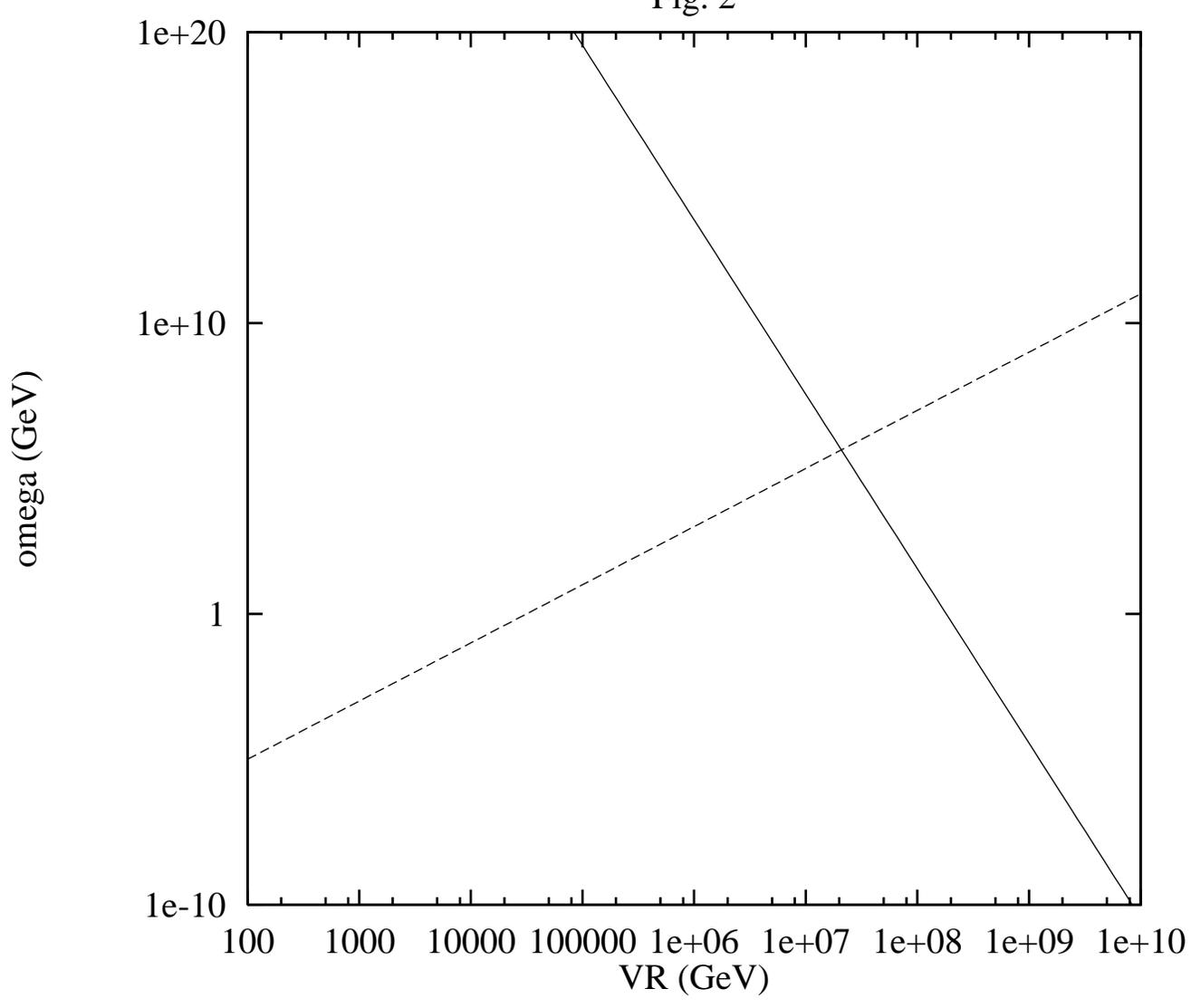
\noindent {\bf Fig. 2}: Constraints on the singlet ($\omega$) and
right-handed doublet ($v_R$) VEVs following from cosmological limit 
\eq{RHO2} (solid line) and from the red giant constraint \eq{red}
(dashed line) for an illustrative choice of the parameter 
$h^2\frac{\sqrt{g_6}}{g_5^3}=10^5$. 
The region below both lines is allowed.

\end{document}